\RequirePackage{lineno}
\documentclass[prd,superscriptaddress,unsortedaddress,twocolumn,showpacs,floatfixamsmath,amssymb]{revtex4}

 \newcommand{\beq}{\begin{equation}}
 \newcommand{\eeq}{\end{equation}}
 \newcommand{\bqa}{\begin{eqnarray}}
 \newcommand{\eqa}{\end{eqnarray}}

\newcommand{\Kz}{\mbox{$K^{0}$}}
\newcommand{\Kzbar}{\mbox{$\overline{K^{0}}$}}

\newcommand{\KL}{\mbox{$K_{L}$}}
\newcommand{\KS}{\mbox{$K_{S}$}}
\newcommand{\Kl}{\mbox{$K_{L}$}}

\newcommand{\epsrat}{\mbox{$\epsilon^{\prime}\!/\epsilon$}}
\newcommand{\reepoe}{\mbox{$Re(\epsrat)$}}

\def\kpi0{K_{L}\to 3\pi^0}
\def\ke3{K_{L}\to\pi^{\pm}e^{\mp}\nu}
\def\km3{K_{L}\to\pi^{\pm}\mu^{\mp}\nu}
\def\k2pi{K_{L} \to \pi^+\pi^-}

\newcommand{\KLpmz}{\mbox{$K_{L}\rightarrow \pi^{+}\pi^{-}\pi^{0}$}}
\newcommand{\Kzzz}{\mbox{$K_L \rightarrow \pi^{0}\pi^{0}\pi^{0}$}}

\newcommand{\KLel}{\mbox{$\Kl\rightarrow \pi^{\pm} e^{\mp}\nu_e$}}

\newcommand{\piz}{\mbox{$\pi^{0}$}}
\newcommand{\eeg}{\mbox{$e^+e^-\gamma$}}

\newcommand{\chisqzz}{\chi^2_{\pi^0}}
\newcommand{\chisqvtx}{\chi^2_{vtx}}

\newcommand{\chisqshape}{\chi^2_{\gamma}}
\newcommand{\dzvtx}{\Delta z_{vtx}}

\newcommand{\sigdz}{\mbox{$\sigma_{\Delta z}$}}

\newcommand{\pgg}{\mbox{$\pi^{0} \rightarrow \gamma\gamma$}}
\newcommand{\pdal}{\mbox{$\pi^{0} \rightarrow e^+e^-\gamma$}}
\newcommand{\pdalg}{\mbox{$\pi^{0} \rightarrow e^+e^-\gamma\gamma$}}
\newcommand{\brdal}{\mbox{$B(\pi^{0} \rightarrow e^+e^-\gamma)/B(\pi^{0} \rightarrow \gamma\gamma)$}}
\newcommand{\Kdal}{\mbox{$K_{L} \rightarrow 3\pi^{0}_{D}$}}
\newcommand{\ee}{\mbox{$e^+e^-$}}
\newcommand{\mee}{m_{\ee}}
\newcommand{\mgg}{\mbox{$m_{\gamma\gamma}$}}
\newcommand{\thrpidal}{\mbox{$3\pi^{0}_{D}$}}
\newcommand{\thrpiz}{\mbox{$3\pi^0$}}
\usepackage{epsfig}
\usepackage{graphicx}
\usepackage{dcolumn}
\usepackage{bm}

\begin{document}

\preprint{draft}

\title{A Measurement of the Branching Ratio of $\piz$ Dalitz Decay using $\Kzzz$ Decays}

\newcommand{\UAz}{University of Arizona, Tucson, Arizona 85721}
\newcommand{\UCLA}{University of California at Los Angeles, Los Angeles,
                    California 90095} 
\newcommand{\Campinas}{Universidade Estadual de Campinas, Campinas, 
                       Brazil 13083-970}
\newcommand{\EFI}{The Enrico Fermi Institute, The University of Chicago, 
                  Chicago, Illinois 60637}
\newcommand{\UB}{University of Colorado, Boulder, Colorado 80309}
\newcommand{\ELM}{Elmhurst College, Elmhurst, Illinois 60126}
\newcommand{\FNAL}{Fermi National Accelerator Laboratory, 
                   Batavia, Illinois 60510}
\newcommand{\Osaka}{Osaka University, Toyonaka, Osaka 560-0043 Japan} 
\newcommand{\Rice}{Rice University, Houston, Texas 77005}
\newcommand{\SaoPaolo}{Universidade de S\~ao Paulo, S\~ao Paulo, Brazil 05315-970}
\newcommand{\UVa}{The Department of Physics and Institute of Nuclear and 
                  Particle Physics, University of Virginia, 
                  Charlottesville, Virginia 22901}
\newcommand{\UW}{University of Wisconsin, Madison, Wisconsin 53706}
\newcommand{\DESY}{DESY, Hamburg, Germany}
\newcommand{\BNL}{Brookhaven National Lab}

\affiliation{\UAz}
\affiliation{\UCLA}
\affiliation{\Campinas}
\affiliation{\EFI}
\affiliation{\UB}
\affiliation{\ELM}
\affiliation{\FNAL}
\affiliation{\Osaka}
\affiliation{\Rice}
\affiliation{\SaoPaolo}
\affiliation{\UVa}
\affiliation{\UW}

\author{E.~Abouzaid}	  \affiliation{\EFI}
\author{M.~Arenton}       \affiliation{\UVa}
\author{A.R.~Barker}      \altaffiliation[Deceased.]{ } \affiliation{\UB}
\author{L.~Bellantoni}    \affiliation{\FNAL}
\author{E.~Blucher}       \affiliation{\EFI}
\author{G.J.~Bock}        \affiliation{\FNAL}
\author{E.~Cheu}          \affiliation{\UAz}
\author{R.~Coleman}       \affiliation{\FNAL}
\author{M.D.~Corcoran}    \altaffiliation[Deceased.]{ }\affiliation{\Rice}
\author{B.~Cox}           \affiliation{\UVa}
\author{A.R.~Erwin}       \altaffiliation[Deceased.]{ }\affiliation{\UW}
\author{C.O.~Escobar}     \affiliation{\Campinas}
\author{A.~Glazov}        
   \altaffiliation[Permanent address ]{\DESY}
   \affiliation{\EFI}
\author{A. Golossanov}     \affiliation{\UVa}
\author{R.A.~Gomes}       \affiliation{\Campinas}
\author{P. Gouffon}       \affiliation{\SaoPaolo}
\author{Y.B.~Hsiung}      \affiliation{\FNAL}
\author{D.A.~Jensen}      \affiliation{\FNAL}
\author{R.~Kessler}       \affiliation{\EFI}
\author{K.~Kotera}	  \affiliation{\Osaka}
\author{A.~Ledovskoy}     \affiliation{\UVa}
\author{P.L.~McBride}     \affiliation{\FNAL}

\author{E.~Monnier}
   \altaffiliation[Permanent address ]{C.P.P. Marseille/C.N.R.S., France}
   \affiliation{\EFI}  

\author{H.~Nguyen}       \affiliation{\FNAL}
\author{R.~Niclasen}     \affiliation{\UB}
\author{D.G.~Phillips~II} \affiliation{\UVa}
\author{E.J.~Ramberg}    \affiliation{\FNAL}
\author{R.E.~Ray}        \affiliation{\FNAL}
\author{M.~Ronquest}     \affiliation{\UVa}
\author{E.~Santos}       \affiliation{\SaoPaolo}
\author{W.~Slater}       \affiliation{\UCLA}
\author{D.~Smith}        \affiliation{\UVa}
\author{N.~Solomey}      \affiliation{\EFI}
\author{E.C.~Swallow}    \altaffiliation[Deceased.]{ }\affiliation{\EFI}\affiliation{\ELM}
\author{P.A.~Toale}      \affiliation{\UB}
\author{R.~Tschirhart}   \affiliation{\FNAL}
\author{Y.W.~Wah}        \affiliation{\EFI}
\author{J.~Wang}         \affiliation{\UAz}
\author{H.B.~White}      \affiliation{\FNAL}
\author{J.~Whitmore}     \affiliation{\FNAL}
\author{M.~J.~Wilking}      \affiliation{\UB}
\author{B.~Winstein}     \altaffiliation[Deceased.]{ } \affiliation{\EFI}
\author{R.~Winston}      \affiliation{\EFI}
\author{E.T.~Worcester}     \altaffiliation[Permanent address ]{\BNL}
\affiliation{\EFI}
\author{T.~Yamanaka}     \affiliation{\Osaka}
\author{E.~D.~Zimmerman} \affiliation{\UB}
\author{R.F.~Zukanovich} \affiliation{\SaoPaolo}

\date{\today}

\begin{abstract}
We present a measurement of $\brdal$, the Dalitz branching ratio,
 using data taken in 1999 by the
E832 KTeV experiment at Fermi National Accelerator Laboratory. We use
neutral pions from fully reconstructed $\KL$ decays in flight; the
measurement is based on $\sim$60 thousand $\Kzzz \rightarrow 
\gamma\gamma ~ \gamma\gamma ~ \ee \gamma$ decays. We normalize to
$\Kzzz \rightarrow 6\gamma$ decays. We find $\brdal$ ($\mee > 15$ MeV/$c^2$)
= $[3.920 \pm
0.016(stat) \pm 0.036(syst)] \times 10^{-3}$. Using the Mikaelian and
Smith prediction for the $\ee$ mass spectrum, 
we correct the result to the full $\ee$ mass range. The
corrected result is $\brdal$ = $[1.1559 \pm 0.0047(stat) \pm
0.0106(syst)]\%$. This result is consistent with previous measurements
and the uncertainty is a factor of three smaller than any previous measurement.
\end{abstract}

\pacs{13.25.Cq}
\maketitle

\section{\label{sec:intro}Introduction}
The neutral pion decays electromagnetically to two photons with a branching ratio of $\sim$99\%. The next most common decay mode, $\pdal,$ was first suggested
by Richard Dalitz in 1951. He calculated the leading order (QED) decay rate 
relative 
to $\pgg$ to be $\brdal = 1.185\%$ \cite{Dalitz}. Radiative corrections to the Dalitz decay
rate have since been calculated to order $\alpha^2$ and predict 
$\brdal = 1.196\%$~\cite{Joseph, Lautrup, Mikaelian, Kampf1, Kampf2, Husek:2015sma}.
A recent calculation using alternative methods reports $\brdal = (1.1978 \pm 0.0006)\%$~\cite{Husek:2018qdx}.
The Dalitz decay branching fraction
has previously been measured to be $\brdal = (1.188 \pm 0.035)\%$\cite{PDG}. The Dalitz decay is used as
a normalization mode for a number of rare kaon and pion decays, and the 
$\sim$3\% uncertainty
in the Dalitz branching ratio measurement is a limiting factor for many of these measurements. 

This
paper reports a new measurement of the Dalitz decay rate using $\Kzzz$ decays in which one of 
the three pions decays to $\eeg$ ($\Kdal$). $\Kzzz$ decays in which
all three pions decay to two photons are used for normalization. 
The measurement is based on 63,693 $\Kdal$
decays collected from June to September 1999 by the KTeV experiment at the Fermi 
National Accelerator Laboratory (FNAL).
In Sec. \ref{sect:expt}, we describe the KTeV beam, experimental apparatus, 
event reconstruction,
and data analysis techniques. In Sec. \ref{sect:ana}, we describe the Dalitz branching
ratio analysis, including corrections to the branching ratio and systematic uncertainties.
Section \ref{sect:result} contains the branching ratio result and crosschecks 
of that result. Section \ref{sect:conclude} provides a comparison to other results
and the new world average.

\section{\label{sect:expt}The KTeV Experiment}
In the KTeV experiment, two neutral kaon beams were produced by a proton
beam incident on a target.
The 800-GeV/c proton beam, provided by the Fermilab Tevatron, had a 53-MHz RF
structure so that the protons arrived in $\sim$1-ns wide ``buckets'' at 19-ns
intervals. The proton extraction cycle was 40-second extractions every 80 seconds.
About half of the data collected in 1999 was at an average intensity of $1.6\times10^{11}$
protons/s with the other half collected at a lower intensity of about $1\times10^{11}$
protons/s
as a systematic cross-check.
The primary purpose of the KTeV experiment was the 
measurement of $\reepoe$~\cite{prd11}. For this reason, a ``regenerator'' 
was placed in one of the beams to produce a source
of $\KS$ decays; this beam is called the regenerator beam and the other beam is called the
vacuum beam. For the Dalitz branching ratio analysis, we use only 
$\KL$ decays from the vacuum beam. A charged
spectrometer was used to measure the momenta and trajectories of charged particles, while
the Cesium Iodide (CsI) calorimeter was used to measure the positions and energies of photons
and electrons. A veto system was used to reject background and a three-level trigger was used
to select events. A detailed Monte Carlo simulation was used to predict the acceptance
difference between the signal and normalization decay modes, and to study background.
The following sections give a brief description of the KTeV detector and reconstruction
techniques; these are described in more detail in \cite{prd03,vus,prd11}.

\subsection{\label{sect:ktevdet}The KTeV Detector}
\begin{figure}
\begin{center}
\epsfig{file=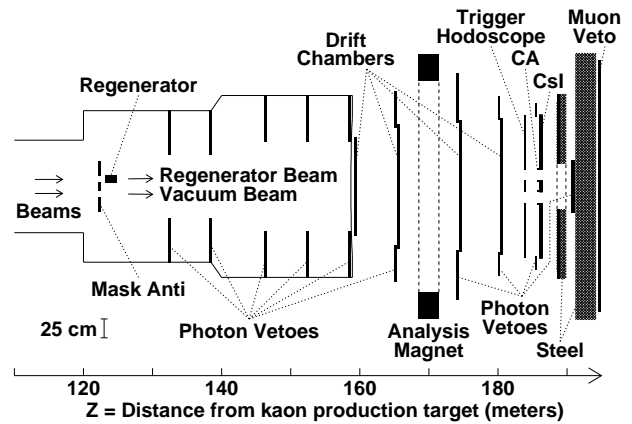,width=\linewidth}
\caption{Schematic of the KTeV detector. Note that the vertical and horizontal
scales are different.}
\label{fig:detector}
\end{center}
\end{figure}

The KTeV kaon beams were produced by an 800 GeV/$c$ proton beam, provided by the FNAL Tevatron,
incident on a beryllium oxide (BeO) target that was about one proton interaction length long.
In the KTeV
coordinate system, the positive $z$-axis points downstream with its origin at the target. 
The two beams were shaped and the non-kaon content was reduced by a beamline of magnets, absorbers,
and collimators. There was an evacuated decay region surrounded by lead-scintillator photon
veto detectors from 90-160 meters downstream of the target. After the vacuum decay region,
there were the charged spectrometer, trigger hodoscope, CsI calorimeter, and muon veto systems.
Figure \ref{fig:detector} is a schematic of the detector.

The charged spectrometer was composed of four drift chambers at $z$=159 m, $z$ = 166 m, $z$ = 175 m,
and $z$ = 181 m, and a dipole analyzing magnet at $z$ = 170 m. Each drift chamber consisted of four
planes of sense wires; two were horizontal and two were vertical. Each sense wire was surrounded
by six field-shaping wires, resulting in a hexagonal cell geometry in each plane. The electron
drift velocity was $\sim 50 \mu$m/ns in the equal-parts argon-ethane gas mixture inside the drift
chambers; this corresponds to a maximum drift time across each cell of 150 ns. The two planes of
sense wires in each view were offset from each other by half a cell to resolve the left-right
ambiguity.
The magnet produced a field which
was uniform to better than 1\% and provided a 0.41 GeV/$c$ momentum kick in the horizontal
plane. The known kaon mass was used to set the momentum scale with $10^{-4}$ precision.

The CsI calorimeter was composed of 3100 pure Cesium Iodide crystals that were each viewed
by a photomultiplier tube. The CsI crystals in the inner region of the calorimeter
were 2.5 $\times$ 2.5 cm$^2$ in the transverse plane and the crystals in the 
outer region were 5.0 $\times$
5.0 cm$^2$. The crystals were all 50 cm (27 radiation lengths) long; 
therefore most of the energy from photons and electrons hitting the CsI calorimeter was measured
by the detector. Two square, carbon-fiber beam holes allowed the beams to pass through the
calorimeter. Momentum analyzed electrons and positrons from $\ke3$ decays were used to calibrate
the CsI energy scale to 0.02\%.

A three-level trigger was used to select events during data collection. Level 1 used fast signals
from the detector, Level 2 was based on processing from custom electronics, and Level 3 was a 
software filter. One of the Level 2 processors was the Hardware Cluster Counter (HCC), which counted
isolated clusters of energy in the CsI calorimeter.
The $\Kdal$ decays for this analysis were selected by a trigger that required
seven or more HCC clusters, while the $\Kzzz$ events for the normalization mode were
selected by a different trigger requiring six or more HCC clusters.

\subsection{\label{sect:ktevana}Event Reconstruction}
Track reconstruction is performed by combining ``hits'' into ``tracks.''
A hit is defined as an analog signal in a drift chamber sense wire that is above
TDC threshold and is in-time with the trigger signal.
The hits in
the two $x$ or $y$ planes of a drift chamber are called a hit-pair. 
For each hit pair, the sum of drift
distances (SOD) should be equal to the cell size, assuming a track that is perpendicular to
the drift chamber and perfect resolution. Hit-pairs are required to have a SOD within 1 mm
of the 6.35 mm nominal cell width after correcting for the incident angle of the particle. 

Track
segments are constructed separately from hit-pairs in the two drift chambers upstream of the magnet
and the two drift chambers downstream of the magnet; these segments are then extrapolated to
the center of the magnet. We require that the extrapolated track segments match to within 6 mm
at the magnet mid-plane. Each particle momentum is determined from the track bend-angle in the
magnet and a map of the magnetic field. If two $x$  and $y$ tracks are found, we extrapolate
both sets of tracks upstream to define an $x-z$ and a $y-z$ vertex. The difference between these
two projections, $\Delta z_{vtx},$ is used to define a vertex $\chi^2$,
\begin{equation}
   \chisqvtx \equiv \left( {\dzvtx}/\sigdz \right)^2 ~,
   \label{eq:chisqvtx}
\end{equation}
where $\sigdz$ is the resolution of $\dzvtx$. A track is required to have $\chisqvtx$ less than 100.
To determine the full particle trajectory,
the $x$ and $y$ tracks are matched to each other based on their projections to the CsI calorimeter.
Each extrapolated track position must match the position of a CsI calorimeter cluster to
within 7 cm.

The energies of photons and electrons are determined by measuring the energy deposited in
the CsI calorimeter by electromagnetic showers. We define a ``cluster'' as a $7 \times 7$
array of small crystals or a $3 \times 3$ array of large crystals. Each cluster is centered
on a ``seed'' crystal which contains the maximum energy deposit among crystals in the cluster.
The energies in all the crystals in the cluster are summed; this sum is then corrected to account
for partial clusters, energy leakage outside the cluster, energy shared between clusters,
and non-uniform detector response.

To reconstruct the decay vertex from clusters in the CsI calorimeter, we group pairs of photons
and determine which pairing produces the most consistent values for the decay vertex. For
each photon pairing, we calculate $d_{12}$, the distance in $z$ between the 
$\piz$ decay vertex and  $z_{\rm CsI}$, the mean shower depth in the CsI crystals.
Using the
pion mass as a constraint, in the 
small angle approximation,
we find the distance from the CsI calorimeter to the vertex for each pair of photons to be 
\beq
d_{12} \approx \frac{\sqrt{E_1E_2}}{m_{\pi^0}}r_{12},
\eeq
where
$r_{12}$ is the transverse distance
between the two photons at the CsI calorimeter.
For each pairing, we compare the calculated distances for each candidate pion.  
The consistency of the reconstructed distances is quantified using the
pairing chi-squared variable ($\chisqzz$) which is based on the reconstructed vertex 
positions and resolutions.
We choose the pairing that gives the minimum value of $\chisqzz$ and require that this
value be less than 75.

Each pion in the signal and normalization modes decays either to two photons or to $\eeg$.
We use $\chisqzz$ to determine which particles come from the same pion. In $\Kdal$ events, 
there are two tracks and seven clusters; two of the clusters are paired to tracks leaving 
five clusters which must
be separated into two pairs of photons from the two $\pgg$ decays and one photon from 
the $\pdal$
decay. In $\Kzzz$ events there are six clusters which must be paired.

The $x$ and $y$ positions of the kaon decay vertex are calculated using the reconstructed
positions and energies of the clusters in the CsI calorimeter. In the case of
tracks from the $\pdal$ decay, the cluster positions are adjusted based on the upstream
track segment, since the cluster's actual position in the calorimeter is the result of
the track bending in the magnet. The CsI cluster energies and the decay vertex position
are used to calculate the invariant mass for the $\thrpidal$ and $\thrpiz$ decays.

\subsection{\label{sect:mc}Monte Carlo Simulation}
We use the KTeV Monte Carlo (MC) simulation to determine the acceptance difference between
the signal and normalization modes and to study background.
The Monte 
Carlo simulates $\Kz$/$\Kzbar$ generation at the BeO target following the
parameterization in \cite{malensek}, propagates the coherent $\Kz$/$\Kzbar$
state through the absorbers and collimators along the beamline to the decay point,
simulates the decay,
traces the decay products through the detector,
and simulates the detector response including the digitization of the 
detector signals and the trigger selection.  The parameters of the detector geometry are 
based both
on data and survey measurements.
Many aspects of the 
tracing and detector response are based on samples of detector responses, 
called ``libraries,'' that are generated with 
GEANT3~\cite{GEANT} simulations.

The effects of accidental activity are included in the simulation by 
overlaying data events from an accidental trigger onto the simulated events.  
The accidental events used in
the simulation are collected concurrently with the signal and normalization data,
so that variations in accidental activity with changes in beam intensity are
simulated by the MC.
After veto requirements are applied, the average accidental energy contained in each
CsI calorimeter cluster is a few MeV and there are about twenty extra in-time
drift chamber hits in each event~\cite{prd03}. 

Simulation of inefficiencies and systematic effects in the drift chambers is
crucial for the Dalitz branching ratio measurement. Individual wire inefficiencies,
high-SOD hit-pairs from delayed hits, accidental hits that obscure hit signals,
and low-SOD pairs from delta rays are all simulated by the MC.

We simulate both real and virtual radiative corrections to the $\pdal$ decay.
QED processes up to
order $\alpha^2$ are included in the simulation; both real corrections, in
which a photon radiates from one of the electrons, and virtual corrections, in which
one-loop terms interfere with the tree-level diagram, are included. 
Real radiative $\pdalg$
events are generated
for $\mgg$ greater than 1 MeV; below this threshold the real radiative process
is indistinguishable in the KTeV detector from the tree-level process. A real
radiative photon above the 1 MeV threshold is generated in about $16\%$ of 
the events.
The virtual corrections are based on \cite{Mikaelian}, which provides numerical
results for radiative corrections over the full kinematic range of the $\ee$ mass
and the energy partition between the electron and positron. A two-dimensional
look-up table binned in the Kroll-Wada $x$ and $y$ variables, which are functions
of the $\ee$ mass and the electron-positron energy partition, respectively, is used to select
the appropriate correction factor.

The Monte Carlo event format is identical to data, and the events are 
reconstructed and analyzed in the same manner as data.
More details of the simulation are available in \cite{vus}.

\section{\label{sect:ana}Branching Ratio Analysis}
To measure the Dalitz branching ratio, we collect $\Kzzz$ events in which 
one of the three
pions undergoes Dalitz decay ($\Kdal$). The normalization mode is 
$\Kzzz$ in which each of the
three pions decays to two photons. The detector acceptance for $\Kdal$ and $\Kzzz$ decays
is quite different, so we use the Monte Carlo simulation to determine the $z$-dependent
acceptance for each mode. We correct this acceptance for data-MC differences related
to tracking inefficiencies and relative trigger differences between the two modes, and
we assign systematic uncertainties associated with these corrections. We
study additional sources of systematic uncertainty on the branching ratio measurement,
such as our simulation of radiative corrections and interactions in the detector material.
In Sec. \ref{sect:dalana}, we describe the selection of signal and normalization events.
In Sec. \ref{sect:dalaccept}, we describe our determination of the relative acceptance of the
two modes, the corrections we make to that acceptance, and the associated systematic
uncertainties. Section \ref{sect:dalsyst} describes
the remaining sources of systematic uncertainty on the branching ratio measurement. A summary
of the systematic uncertainties is given in Sec. \ref{sect:systsummary}. 

\subsection{\label{sect:dalana}Event Selection}
The $\Kdal$ signal and $\Kzzz$ normalization mode events are selected by separate triggers 
that require 
seven or more clusters and six or more clusters, respectively, in the CsI calorimeter. 
To ensure
consistency between the signal and normalization modes, any runs or spills that do not
contain both types of triggers are excluded.

We apply a number of selection criteria to the $\Kdal$ and $\Kzzz$ events. 
These
cuts are chosen to avoid event topologies that have poor reconstruction efficiencies
or that are difficult to simulate, to reduce backgrounds, and to define the 
acceptance. We keep the selection criteria for signal and normalization events as
similar as possible so that any associated systematic uncertainties cancel in
the ratio of the two modes. For those requirements associated with the tracks in
the $\pdal$ decay, we vary each requirement in both data and MC events to verify that the
data are well-simulated by the MC in the region of that cut.

We eliminate signal events with more than two tracks, and both signal
and normalization events with extra clusters in the CsI calorimeter. In both modes,
we require that the reconstructed invariant mass be within 7 MeV/$c^2$ of the known
kaon mass, that the reconstructed $z$ vertex position be between 123 and 158 m, and that
the reconstructed kaon energy be between 40 and 160 GeV. 

To avoid events in which the
CsI calorimeter clusters are difficult to reconstruct and simulate, we place a number
of requirements on the CsI clusters for both data and Monte Carlo. 
The minimum cluster energy in an event must be
greater than 3 GeV, and the minimum distance between clusters must be greater than 7.5 cm.
The ``ring number'' variable~\cite{prd11}, which describes the distance between the center of energy and
the nearest beamhhole, must be less than 110 cm$^2$. Events in which the kaon decay
occurred within one of the neutral beams should have a ring number less than 86.5 cm$^2$~\cite{prd11}.
The $\chisqshape$ variable, which
describes how close a cluster's transverse energy distribution is to the distribution
expected for a photon shower
\endnote{The $\chisqshape$ variable is not normalized to be a true chi-squared.}, 
must be less than 100. We remove events in which one of the clusters has its seed crystal
in the innermost or outermost ring of crystals in the CsI calorimeter. These criteria are
identical for both the signal and normalization modes.

A number of additional requirements are placed on the $\Kdal$ signal mode to select
 events in which the tracks are well-reconstructed and simulated. We require that the
reconstructed $\eeg$ invariant mass be within 20 MeV/$c^2$ of the known $\piz$ mass. The
resolution of the $\eeg$ invariant mass is $\sim 1.5$ MeV/$c^2$. The minimum track
momentum must be greater than 4.0 GeV/$c$, and the ratio of measured energy in the CsI
cluster associated with the track to the measured track momentum, E/$p$, must be greater 
than 0.9.

We remove events in which a track passes too close to the edge of a detector. 
These
fiducial cuts reduce the sensitivity of the measurement to our knowledge of the physical
size and location of these detectors. The tracks must be more than 2-3 mm away from 
inner edges of the veto detectors 
and trigger hodoscope that surround the neutral beams and at least 
2.9 cm away from the outer edge of the CsI calorimeter.

We remove events in which one of the electrons emits a bremsstrahlung photon as it
bends in the magnet. The bremsstrahlung photon is typically emitted parallel to
the direction of the electron prior to bending in the magnet.
To identify these events, we project the upstream segment of each
track to the the CsI calorimeter and identify the closest photon cluster. The distance
between the track projection and the position of the nearest cluster is called the
``brem-$\gamma$ distance''; we require this quantity to be greater than 1 cm.

A critical requirement on $\Kdal$ events is that the $\ee$ tracks be separated
by more than three drift chamber cells in the two upstream drift chambers. This
corresponds to a distance requirement of more than $\sim$2 cm. The MC simulation
of tracks in the same or neighboring cells, complicated by other effects such as
accidental hits, delta rays, and high-SOD pairs, is difficult; this cell separation
requirement is necessary to ensure that the tracking efficiency is well-modeled.

We also require that the reconstructed $\ee$ mass be greater than 15 MeV/$c^2$;
this requirement is related to the cell separation requirement because small values
of $\ee$ mass correspond to close tracks. Since the cell separation requirement 
removes most events with a reconstructed $\ee$ mass less than 10 MeV/$c^2$, the 
analysis requirement on $\ee$ mass cleanly defines the kinematic region of our
measurement by excluding the region where the acceptance is very small. Additionally,
the low $\ee$ mass region is more sensitive to real and virtual radiative 
corrections than the region above 15 MeV/$c^2$. Removing the low mass region
reduces the sensitivity of this measurement to theoretical predictions and allows
the measurement to be updated in the future when new calculations of radiative
corrections at low $\ee$ mass are available.

\subsection{\label{sect:dalaccept}Acceptance and Acceptance Corrections}
The detector acceptance for $\Kdal$ and $\Kzzz$ decays is the ratio
of simulated events passing all reconstruction criteria to the total number of events
simulated. Monte Carlo events are generated in a larger kinematic range than will be
accepted by the analysis, so that the possibility of event migration across selection
criteria boundaries is treated correctly. The MC simulation includes kaon decays in the range
110 m $ < z <$ 161 m and 35 GeV $ < E < $ 165 GeV. Dalitz decays are simulated
for all possible $\ee$ masses.  

We simulate $\sim$200 million $\Kdal$ decays and $\sim$300 million $\Kzzz$ decays.
Out of these, the fraction of Dalitz events accepted by the analysis is 
$(1.1714 \pm 0.0023)\times 10^{-3}$. The fraction of events accepted for $\Kzzz$ decays is 
$(3.7853 \pm 0.0010)\times 10^{-2}$. The quoted uncertainties are from MC statistics only.
There are two differences between the data and MC
that have a significant effect on the acceptance calculation: the simulation
of tracking efficiencies in the signal mode, and the simulation of the relative trigger
efficiencies between the signal and normalization modes.
We correct the acceptance determined from Monte Carlo
for these known differences between data and the Monte Carlo simulation.
The following sections describe how we determine the appropriate acceptance correction
for each of these discrepancies.

\subsubsection{\label{sect:trkcorr}Acceptance Correction for Tracking Efficiencies}
In Dalitz decay, the angle between the electron and positron tends to be small, so
the track separation in the two upstream drift chambers is small. It is difficult
to simulate tracking inefficiency for close tracks, so we must correct for data-MC 
differences in tracking inefficiency.

We use an independent sample of
$\KLpmz$ decays to measure the tracking efficiency for two charged particles in the KTeV
detector because these events can be reconstructed without full tracking. We reconstruct the decay 
vertex position of 
the $\pgg$ decay using the positions and energies of clusters in the CsI calorimeter. The two
charged pions are associated with hadronic showers in the CsI calorimeter; these are differentiated
from the photon clusters using the $\chisqshape$ variable. We then use this sample to measure
the single-track and two-track inefficiencies in data and Monte Carlo as described below.

To measure the single-track inefficiency, we require one of the hadronic clusters in the CsI
calorimeter to match a fully-reconstructed track. There are two possible kinematic solutions
for the second track; the position of the second hadronic cluster resolves the ambiguity. The
single-track inefficiency, $\eta_1$, is half the ratio of events with a missing track to the 
total number
of events; the factor of two is included since either of the two tracks could be lost.

To measure the probability of failing to reconstruct both tracks, $\eta_0$,
we select events in which there are two track segments in either the upstream or
downstream pair of drift chambers, but no complete tracks are reconstructed. The ratio of these
events to the total is the two-track inefficiency.

These tracking efficiency measurements are performed in both data and Monte Carlo samples at
two different beam intensities, where the average intensities in the two samples differ by
about a factor
of two. The measured inefficiencies are summarized in 
Table \ref{tb:trkineff}.
The data-MC differences for medium and high intensity data are averaged to apply a correction of 
-0.68$\%$ to the acceptance calculation. 

\begin{table}[!htpb]
\caption{Tracking inefficiencies in $\KLpmz$ data and Monte Carlo, for two different
beam intensities. The correction applied to the acceptance is the difference between
the total data inefficiency and the total MC inefficiency.}
\centering
\begin{tabular}{|l|c|c|}
\hline 
   & \multicolumn{2}{c|}{Tracking Inefficiency} \\
   & Medium Intensity & High Intensity \\ \hline \hline
Data & & \\
2$\eta_1$ & 3.48\% & 4.90\% \\ 
$\eta_0$ & 0.19\% & 0.21\% \\
Total & 3.67\% & 5.11\% \\ \hline
Monte Carlo & & \\
2$\eta_1$ & 2.97\% & 4.31\% \\ 
$\eta_0$ & 0.05\% & 0.09\% \\
Total & 3.02\% & 4.40\% \\ \hline
Correction & 0.65\% & 0.72\% \\ \hline
\end{tabular}
\label{tb:trkineff}
\end{table}

Most of the measured tracking inefficiency is due to accidental activity.
To demonstrate this, we perform the tracking efficiency
measurements on Monte Carlo events that do not include accidental
overlays and find that the measured inefficiencies in this sample are much smaller than in the nominal
Monte Carlo. We find that $\sim$90$\%$ of
the measured tracking inefficiency results from accidentals.
The impact of accidental activity is well-modeled in the MC and the observed
data-MC discrepancy is largely independent of intensity, as seen in Table~\ref{tb:trkineff}.
We conclude that much of the -0.68\% correction to the efficiency can not be attributed to accidental
activity.
Since this discrepancy is unexplained, we assign a systematic uncertainty on the Dalitz
branching ratio measurement that is equal to the full size of the correction to the efficiency.

\subsubsection{\label{sect:trigcorr}Acceptance Correction for Relative Trigger Efficiencies} 
The $\Kdal$ signal and $\Kzzz$ normalization events are selected with different triggers.
To measure the relative inefficiencies between the two triggers,
we use a rescaled sample of events from the normalization mode trigger that do not have any 
Level 3 requirements
applied. We apply the $\Kdal$ reconstruction algorithm and selection criteria to this sample and
search for events that would be included in the Dalitz analysis but are not included in the sample
selected by the Dalitz trigger.
All requirements are the same as in the primary analysis except that the cell separation cut is
removed to increase statistics. In this sample of 716 events, we find one event that 
passes
all other Dalitz selection criteria but is not included in the Dalitz sample. The same study is
performed on Monte Carlo events with no measurable relative trigger inefficiency found.
This data-Monte Carlo difference in trigger inefficiency
of 0.14$\%$ is applied as a correction to the acceptance. 
We assign a systematic uncertainty on the Dalitz
branching ratio measurement that is equal to the size of the correction.

We also measure the absolute inefficiency of the trigger used to select $\Kzzz$ decays for
the normalization sample. We study data from a minimum bias trigger and search for events 
that would be accepted by the $\Kzzz$ analysis but were not selected by the $\Kzzz$
trigger. We select a sample of $\sim$500,000 $\Kzzz$ decays from the minimum bias trigger
and measure the trigger inefficiency to be $(0.0042 \pm 0.0010)\%$. There is no trigger
inefficiency simulated in the Monte Carlo, so the full inefficiency is a data-MC bias.
We do not correct for this small inefficiency. We apply the standard
KTeV procedure for setting systematic uncertainties that includes the statistical precision
of the study~\cite{prd03}; we find a systematic uncertainty in the Dalitz branching ratio
measurement of 0.0047$\%$.

\begin{figure}
\begin{center}
\epsfig{file=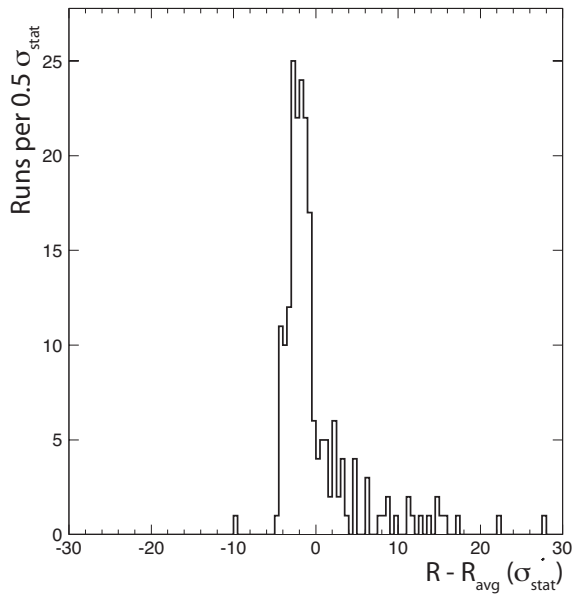,width=\linewidth}
\caption{Distribution of difference from the average value for ratio
of number of events before to number of events after hardware prescale
is applied, in units of statistical standard deviations.}
\label{fig:prescalesyst}
\end{center}
\end{figure}

A prescale of 2 at the hardware level and of 5/2 at the software level
is applied to the $\Kzzz$ trigger; there is no prescale applied to the $\Kdal$ sample.
Any deviation of the prescale from the nominal values will produce a bias in the branching
ratio measurement. The software prescale has no inaccuracy. We study the hardware prescale
accuracy
using scaler counts of the number of events before and after the hardware prescale was
applied during data collection. For each individual run and for all runs combined,
we calculate the ratio of events, R = N$_f$/N$_i$, after prescaling (N$_F$) to before 
prescaling (N$_i$),
and the statistical uncertainty of each ratio. The average ratio is 
R$_{avg}$ = $0.500044 \pm 0.000003$.
In Fig. \ref{fig:prescalesyst},
we plot the number of statistical sigma from average for each run. We find a 
number of runs in which the measured prescale is significantly different from the
nominal value;
this indicates a small,
intermittent defect in the prescale electronics. 
As $\sim$90\% of runs have a discrepancy of less than 5$\sigma$, we assign an 
uncertainty of 5 times the statistical error on the total ratio. We 
therefore determine the prescale ratio
to be 0.500044 $\pm$ 0.000016, which corresponds to systematic uncertainty in
the branching ratio measurement of $<0.01\%$.

\subsection{\label{sect:dalsyst}Other Systematic Uncertainties}
This section contains descriptions of how we assign the remaining
systematic uncertainties for the measurement of $\brdal$. All of the
systematic uncertainties are summarized in Sec. \ref{sect:systsummary}.

\subsubsection{Radiative Corrections}
\begin{figure}
\begin{center}
\epsfig{file=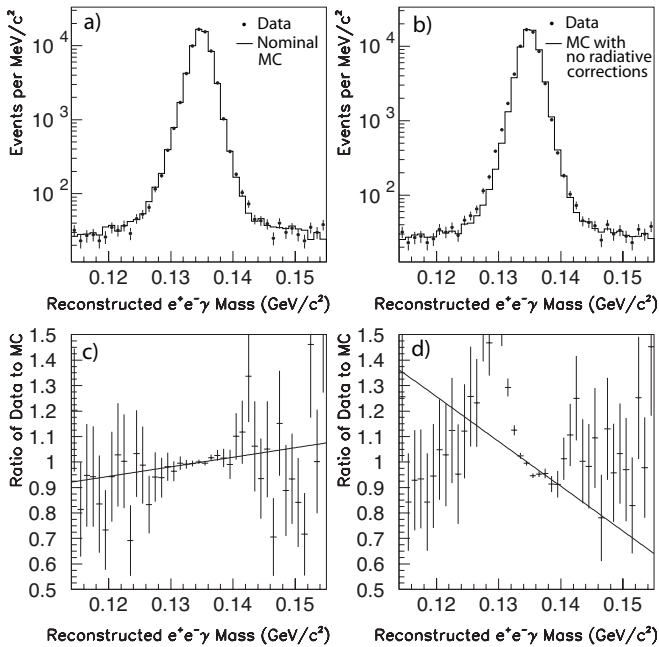,width=\linewidth}
\caption{Reconstructed $\eeg$ mass for data (dots) and 
Monte Carlo (histogram). a) Data and nominal MC. b) Data and MC
with no radiative corrections. c) Data/MC ratio for nominal MC.
d) Data/MC ratio for MC with no radiative corrections. All nominal
selection criteria have been applied
except for the $\eeg$ mass requirement.}
\label{fig:eegmasscomp}
\end{center}
\end{figure}

\begin{figure}
\begin{center}
\epsfig{file=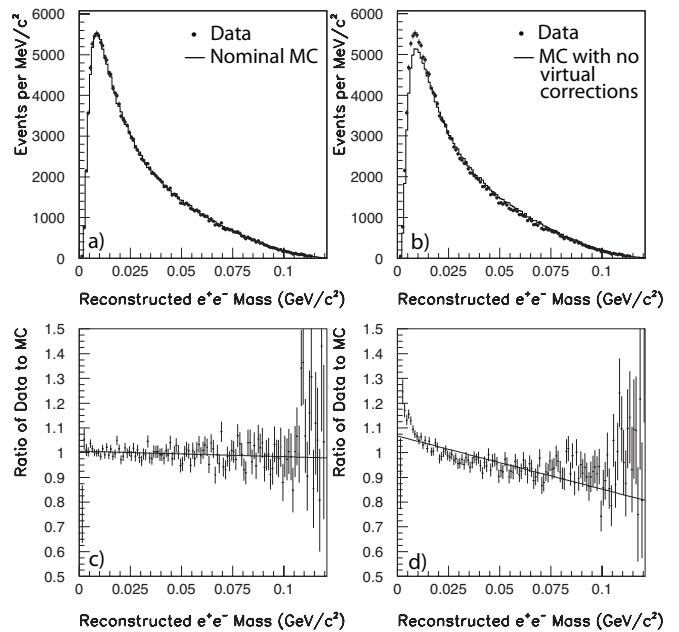,width=\linewidth}
\caption{Reconstructed $\ee$ mass for data (dots) and 
Monte Carlo (histogram). a) Data and nominal MC. b) Data and MC with
no virtual radiative corrections. c) Data/MC ratio for nominal MC. d) Data/MC
ratio for MC with no virtual radiative corrections.
All nominal selection criteria have 
been applied
except for the $\ee$ mass requirement and the cell separation requirement.
These plots include events with minimum cell separation values of one
or greater.}
\label{fig:eemasscomp}
\end{center}
\end{figure}

As described in Sec. \ref{sect:mc},
the Monte Carlo simulation of the $\pdal$ decay includes radiative
corrections to second order in $\alpha_{EM}$. 
The reconstructed $\eeg$ mass distribution is sensitive to the 
real corrections,
while the reconstructed $\ee$ mass distribution
is sensitive to the virtual
corrections; data-MC comparisons for these quantities are shown in
Figs. \ref{fig:eegmasscomp} and \ref{fig:eemasscomp}.
Since the QED calculations
that are used by the simulation are well-understood and data-MC 
comparisons indicate that the data are well-described by the simulation,
we assign a systematic uncertainty on the Dalitz branching ratio due
only to higher-order corrections that are not simulated. The acceptance 
change between Monte Carlo with no radiative corrections and the nominal
MC with second order corrections
is -5.43\%. We assume that adding the
next order of corrections would cause the same percentage change in
acceptance; we therefore take 5.43\% of 5.43\%, or 0.29\%, to be
the systematic uncertainty in the branching ratio measurement due to
higher-order radiative corrections.  

\subsubsection{Detector Material}
The Monte Carlo simulation includes bremsstrahlung radiation in
the 0.018 radiation lengths of detector material located upstream
of the final drift chamber. The simulation of bremsstrahlung in
the nominal Monte Carlo changes the 
signal mode acceptance by -4.66\% relative to a MC with no bremsstrahlung
simulation. The amount of detector material is known
to about 10\%, so we assign a systematic uncertainty on
the branching ratio measurement of 0.47\%. 

\subsubsection{Accidental Activity}
In addition to being the primary source of tracking inefficiency, accidental 
activity can affect the branching ratio measurement by
adding extra tracks or CsI calorimeter clusters to an event.
Accidentals affect the signal and normalization
modes differently because tracks are
present only in the signal mode, and because the normalization mode has one
more photon in the final state than the signal mode. The presence of
accidental overlays in the Monte Carlo simulation decreases the acceptance
relative to a simulation with no accidentals by 37\% for the signal mode
and by 32\% for the normalization mode. These changes largely cancel
in the branching ratio measurement; the change in the branching ratio
from the simulation of accidental activity is 3.96\%. We estimate the
uncertainty in the branching ratio analysis from accidental activity by
combining Monte Carlo samples as described in the following paragraphs.

\begin{figure}
\begin{center}
\epsfig{file=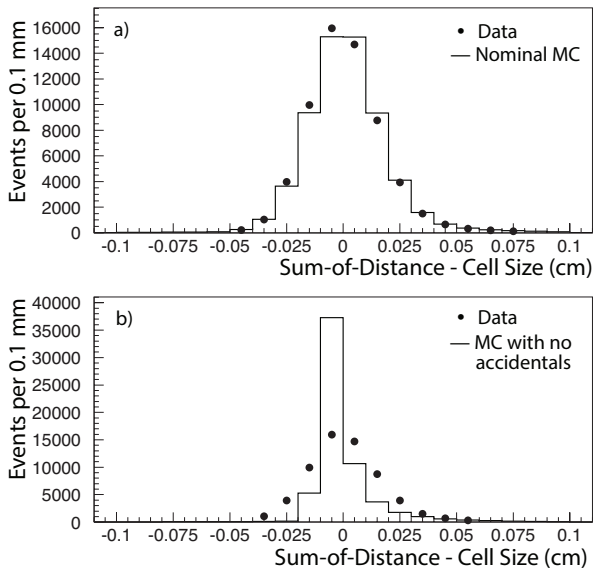,width=\linewidth}
\caption{Deviation of sum-of-distance (SOD) from nominal value of 6.35 mm for
data (dots) and Monte Carlo (histogram). Note the different vertical scales for
the two plots. a) Data and nominal MC. b) Data and MC
with no simulation of accidental activity. }
\label{fig:accoverlay}
\end{center}
\end{figure}

To estimate the sensitivity of the branching ratio measurement to
extra tracks from accidentals, we study events with low sum-of-distances
(SOD) values because these events come almost entirely from accidentals.
As shown in Fig. \ref{fig:accoverlay}, the fraction of events with low 
SODs is well-modeled by the Monte Carlo
simulation; the ``low-SOD fraction,'' defined as the fraction of events with
SOD less than -0.2 mm, in the nominal MC is within 2.5 sigma 
of the fraction in data. To quantify the sensitivity of the measurement
to our simulation of accidentals, we create Monte Carlo samples in
which some fraction of events do not contain accidental overlays and
find the level at which we can detect a data-MC discrepancy. We find that
a Monte Carlo sample composed of 97\% nominal MC and 3\% MC with no
accidentals produces a low-SOD fraction that is significantly different
from the data. Since no significant data-MC difference in the low-SOD fraction
is observed in the
nominal data-MC comparison, we conclude that the effect of accidentals on
tracks is modeled to within 3\%. We therefore
assign a systematic uncertainty on the Dalitz
branching ratio equal to 3\% of 3.96\%, or 0.12\%, due to the
simulation of extra tracks from accidentals.

Accidental activity results in extra ``software
clusters'' in the CsI calorimeter. These extra clusters are found in the
reconstruction but are not energetic enough to be detected by the
trigger during data acquisition. The distribution of software
clusters is reasonably well-modeled by the Monte Carlo simulation;
the fraction of events with no extra software clusters differs by
3.3 sigma between data and Monte Carlo. A Monte Carlo sample that
is composed of 99\% nominal MC and 1\% MC with no accidentals produces 
a data-MC mismatch in the fraction of extra software clusters
that is larger than the observed data-MC discrepancy. We therefore
conclude that the effect of accidentals on photon clusters in the signal
mode is modeled to better than 1\%, and assign a systematic uncertainty on
the Dalitz branching fraction equal to 1\% of 3.96\%, or 0.04\%, due to the simulation
of extra calorimeter clusters from accidentals.

The total uncertainty in the Dalitz branching ratio associated with
the simulation of extra tracks and clusters from accidental activity is 
found by combining the uncertainties from extra tracks and extra clusters 
in quadrature.
The resulting systematic uncertainty in the branching ratio is 0.13\%.

\subsubsection{Form Factor}
The amplitude for the $\pdal$ decay contains a form factor at the 
$\pi^0\gamma\gamma$ vertex. The form factor is approximated by 
$f(x) \approx (1 + ax)$, where $x = (m_{e^+e^-}/m_{\pi})^2$ and
$a$ is the $\pi^0$ slope parameter. The nominal value of the slope
parameter is $a = 0.032 \pm 0.004$~\cite{PDG}\footnote{Since the analysis
  described here was performed, a new measurement of the slope parameter
  has been made by NA62\cite{TheNA62:2016fhr}. The change to the world
  average of the form factor is +0.003 with respect to the value used
  in this analysis; that difference is covered by the systematic uncertainty
  quoted here.}. To determine the
sensitivity of the Dalitz branching ratio to the value of the form factor
used in the Monte Carlo simulation, we measure the change in acceptance,
relative to the nominal MC,
for MC samples with values of $a$ 
that are 8 sigma above and below the nominal
value. The acceptance changes by $(0.388 \pm 0.274)\%$ and 
$(0.155 \pm 0.274)\%$, respectively. Dividing the larger of these
two acceptance changes by 8 results in a one-sigma systematic uncertainty in
the Dalitz branching ratio of 0.06\%.

\subsubsection{Selection Criteria}
We study the systematic uncertainty associated with the analysis
selection criteria by varying each selection requirement and finding
the associated change in the branching ratio measurement. We find
that variations in selection criteria that are common to the signal 
and acceptance modes cancel in the branching fraction as expected.
Selection criteria that are unique to the signal mode also produce 
no significant change in the branching fraction when varied. Since
we find no significant change in the Dalitz branching fraction when
varying any of the selection criteria, we assign no additional systematic uncertainty.

We require the reconstructed $\ee$ mass to be greater than 15 MeV/$c^2$.
Therefore, any disagreement between data and Monte Carlo in the $\ee$ mass 
scale will produce an uncertainty in the branching ratio measurement. We
estimate our sensitivity to the $\ee$ mass scale by varying the scale and
checking the data-MC comparison. We find that a 0.5\% shift in the $\ee$
mass scale results in a detectable data-MC disagreement. Since we
see no significant data-MC disagreement in the nominal analysis, the
$\ee$ mass scale must match to within 0.5\%; the corresponding uncertainty
in the branching ratio measurement is 0.06\%.

We require CsI calorimeter clusters to have a transverse energy distribution
similar to one that is expected for a photon by applying a cut on 
the $\chisqshape$ variable. Since the normalization mode contains one
more photon than the signal mode, any photon inefficiency resulting from
this requirement is a source of systematic uncertainty. We make use of the 
result
from a previous analysis measuring B($\Kzzz$)/B($\KLel$)~\cite{vus},
which has six more photons in the signal mode than in the normalization mode
and found a change of 0.05\% when removing the $\chisqshape$
requirement. We scale this result by 1/6 to estimate the systematic
uncertainty associated with the one-photon difference between the
signal and normalization modes in this analysis; we assign a
systematic uncertainty associated with photon inefficiency of 0.01\%.

\subsubsection{Background}
The primary background to the $\Kdal$ decay is $\Kzzz$ decay in which
one photon converts to an $\ee$ pair at the vacuum window. Using a
large sample of simulated $\Kzzz$ decays, we find that 0.0003\%
of these events are accepted by the $\Kdal$ analysis. This background
is negligible, so we do not subtract it; we take the 0.0003\% background 
rate as a systematic
uncertainty on the branching ratio measurement.

\subsection{\label{sect:systsummary}Summary of Systematic Uncertainties}
Table \ref{table:systsummary} contains a summary of the systematic uncertainties for
the measurement of $\brdal$. Most of the sources of error are related to uncertainty
in the Monte Carlo simulation of the relative acceptance between the two decay modes.
The largest source of uncertainty is from differences in the tracking 
efficiency between
data and Monte Carlo. The total systematic uncertainty on the Dalitz branching ratio
measurement is 0.92\%.

\begin{table}[!htbp]
\caption{Summary of systematic uncertainties in the $\brdal$ branching ratio.}
\centering
\begin{tabular}{|l|c|}
\hline
Source of Uncertainty & Level of Uncertainty \\ \hline \hline
Tracking Inefficiency         &  0.68\% \\
Relative Trigger Inefficiency &  0.14\% \\
$\Kzzz$ Trigger Inefficiency  & $<$0.01\% \\
$\Kzzz$ Trigger Prescale      & $<$0.01\% \\
Radiative Corrections         &  0.29\% \\
Detector Material             &  0.47\% \\
Extra Tracks and Clusters     &  0.13\% \\
Form Factor                   &  0.06\% \\
$\ee$ Mass Scale              &  0.06\% \\
Photon Inefficiency           &  0.01\% \\
Background                    & $<$0.01\% \\
Monte Carlo Statistics        &  0.20\% \\ \hline
Total                         &  0.92\% \\ \hline
\end{tabular}
\label{table:systsummary}
\end{table}

\section{\label{sect:result}Result and Crosschecks}
We find 63,693 $\Kdal$ decays with an acceptance of 0.12\% and
3,529,065 $\Kzzz$ decays with an acceptance of 3.79\%. We scale the
$\Kdal$ acceptance by a factor of 3 since any of the three pions
could undergo Dalitz decay. We scale the $\Kzzz$ acceptance
by 50 to account for the prescale applied during data collection. 
The final result is
\bqa
\lefteqn{\frac{B(\pi^{0} \rightarrow e^+e^-\gamma)}{B(\pi^{0} \rightarrow \gamma\gamma)} 
,\mee > 15~\mbox{MeV/$c^2$} = }\nonumber \\
& [3.920 \pm 0.016(stat) \pm 0.036(syst)] \times 10^{-3}.
\label{eq:resultgt15}
\eqa

\begin{figure}
\begin{center}
\epsfig{file=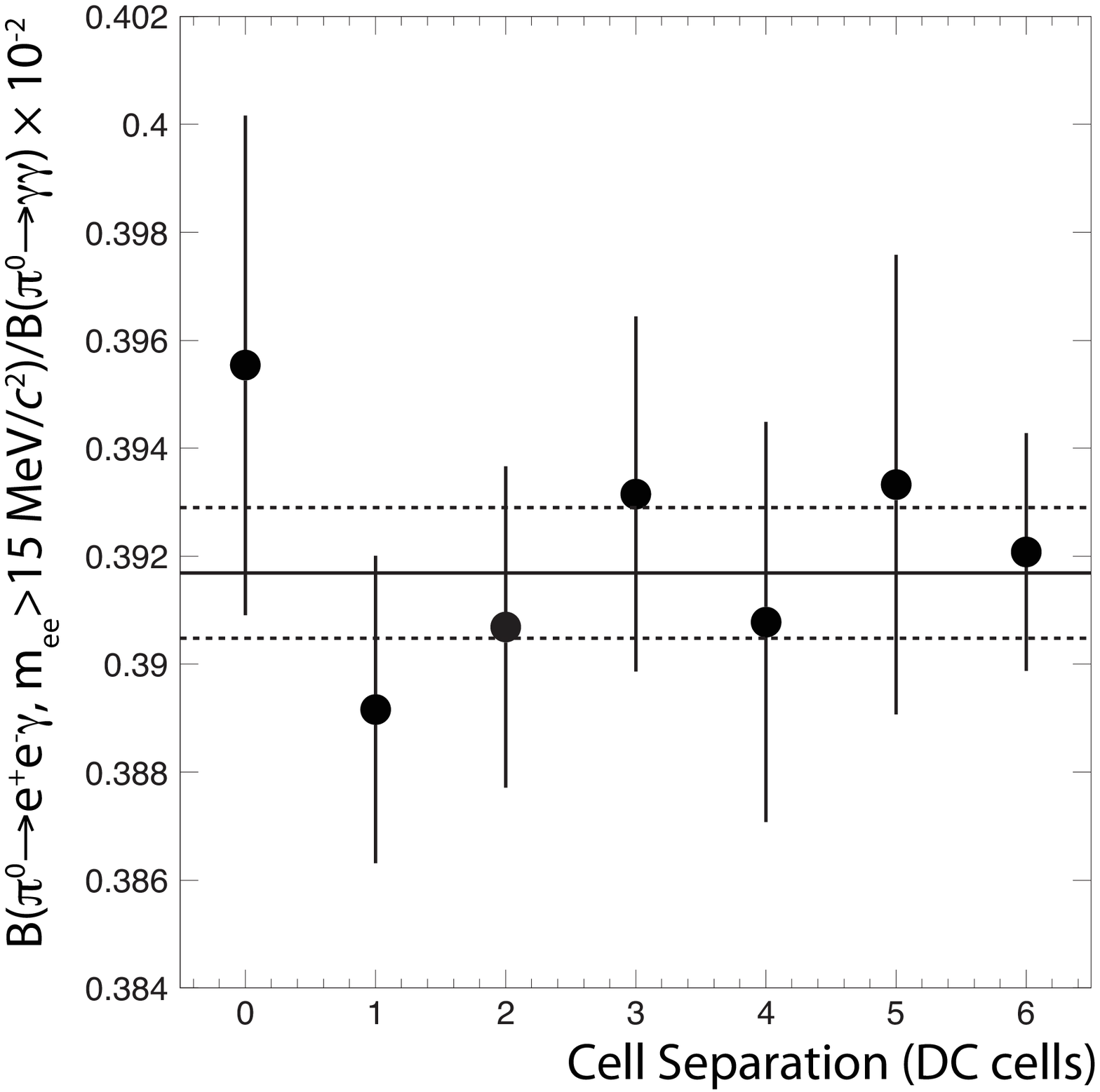,width=0.9\linewidth}
\caption{The $\brdal$ measurement versus cell separation. The Dalitz events
that contribute to the result in each of the first six bins have a minimum cell
separation equal to the bin number. The last bin includes events with minimum
cell separation greater than or equal to six. The error bars represent the independent
statistical uncertainty in each bin. The solid line is the weighted average and the
dashed horizontal lines indicate the statistical uncertainty on the weighted average.}
\label{fig:cellsepvar}
\end{center}
\end{figure}

We check the consistency of the $\brdal$ branching ratio measurement by
comparing the result among subsets of the data. We separate the data into
groups by cell separation, $\ee$ mass, beam intensity, time, whether the
tracks bend toward or away from each other, and the two polarities
of the analysis magnet. In each case, the results in the sub-samples agree
with each other and with the result from the full data sample. Figure 
\ref{fig:cellsepvar} shows the $\brdal$ ($\mee > 15$ MeV/$c^2$)
 result as a function of minimum
 cell separation with the track separation requirement of three cells removed.
 A constant fit to these points has a probability of 92\%;
the result is stable as a function of minimum cell separation.   

We correct the result in Eq. \ref{eq:resultgt15}, which is valid for $\ee$ masses greater than
 15 MeV/$c^2$, to
the full mass range using a calculation of the $\ee$ mass spectrum from Mikaelian and
Smith~\cite{Mikaelian}. We find that 33.9128\% of Dalitz decays occur above the 15 MeV/$c^2$
$\ee$ mass cutoff applied in this analysis. The corrected result, valid over the full $\ee$
mass range, is
\bqa
\lefteqn{\frac{B(\pi^{0} \rightarrow e^+e^-\gamma)}{B(\pi^{0} \rightarrow \gamma\gamma)} =} \nonumber \\
&[1.1559 \pm 0.0047(stat) \pm 0.0106(syst)]\%.  
\label{eq:resultfull}
\eqa

\section{\label{sect:conclude}Conclusion}
\begin{figure}
\begin{center}
\epsfig{file=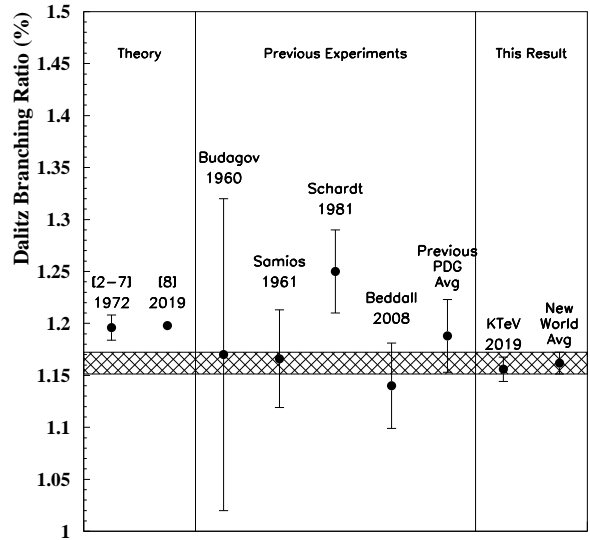,width=\linewidth}
\caption{Comparison of the $\brdal$ branching ratio result to previous
  theoretical~\cite{Joseph, Lautrup, Mikaelian, Kampf1, Kampf2, Husek:2015sma, Husek:2018qdx}
  and experimental~\cite{Budagov,Samios,
Schardt,
Beddall,PDG} results. A new world average for experimental results is also shown.
The hatched region shows the area within one sigma of the new world average.}
\label{fig:resultcomp}
\end{center}
\end{figure}

We have measured the Dalitz decay branching ratio, $\brdal$, for $\ee$ masses greater than
15 MeV/$c^2$ using $\Kdal$ and $\Kzzz$ decays. Correcting to the full $\ee$ mass range, we
find
\beq
\frac{B(\pi^{0} \rightarrow e^+e^-\gamma)}{B(\pi^{0} \rightarrow \gamma\gamma)} = (1.1559 \pm 0.0116)\%.  
\label{eq:result}
\eeq
Figure \ref{fig:resultcomp} is a comparison of this result to the theoretical calculation 
and previous experimental results. This result agrees with the 1972 theoretical calculation 
at the
2.4 sigma level, where a 1\% uncertainty on the calculation has been assumed based on discussion in
\cite{Mikaelian}.
The discrepancy with \cite{Husek:2018qdx} is 3.6 sigma.
The uncertainty in this measurement is at least a factor of three smaller than the individual uncertainties on all previous 
measurements and the uncertainty on the previous PDG average~\cite{PDG}. 
We combine this result with the four
previous measurements to find the new world average is $\brdal = (1.1619 \pm 0.0105)\%$.

\begin{acknowledgments}
We gratefully acknowledge the support and effort of the Fermilab
staff and the technical staffs of the participating institutions for
their vital contributions.  This work was supported in part by the U.S.
Department of Energy, The National Science Foundation, The Ministry of
Education and Science of Japan,
Funda\c c\~ao de Amparo a Pesquisa do Estado de S\~ao Paulo-FAPESP,
Conselho Nacional de Desenvolvimento Cientifico e Tecnologico-CNPq and
CAPES-Ministerio Educa\c c\~ao.

\end{acknowledgments}

\bibliography{dalitz}

\end{document}